# Aqueous Preparation of CsPbBr$_3$ Perovskite Nanocrystals Under Ambient Conditions

*Zhaoyi Du, Jiewen Wei, Ding Ding, Martina Rimmele, Yueyao Dong, Weitao Qian, Davide Nodari, Francesco Furlan, Edoardo Angela, George Morgan, Peter Akinshin, William Rodriguez Kazeem, Gwilherm Kerherve, Adam V. Marsh, Martin Heeney,*
*Thomas J. Macdonald, Saif A. Haque, Nicola Gasparini, David J. Payne*
*and Martyn A. McLachlan\**

Z. Du, W. Qian, E. Angela, W. Kazeem, G. Kerherve, D. J. Payne, M. A. McLachlan
Department of Materials
Imperial College London
London, W12 0BZ, UK
E-mail: martyn.mclachlan@imperial.ac.uk

J. Wei, D. Ding, M. Rimmele, D. Nodari, F. Furlan, S. A. Haque, N. Gasparini
Department of Chemistry
Imperial College London
London, W12 0BZ, UK

Y. Dong, T. J. Macdonald
Department of Electronic and Electrical Engineering
University College London
London, WC1E 6BT, UK

A. V. Marsh, M. Heeney
King Abdullah University of Science & Technology (KAUST)
Thuwal, 23955-6900, Saudi Arabia

G. Morgan, P. Akinshin
Now at Department of Physics
University of Oxford
Oxford OX1 3PU, UK

**Abstract**

Metal halide perovskites (MHPs) have had a profound impact on numerous emerging optoelectronic technologies, achieving performance metrics that rival or exceed incumbent materials. This impact is underpinned by the exceptional properties of MHPs, including tuneable band gaps, high absorption coefficients, long carrier diffusion lengths and combined with uncomplicated synthesis methods. However, current MHP production relies on the toxic solvents, which pose significant environmental and health risks. Moreover, these methods often require complex multi component solvent systems and thermal processing to achieve the desired material phases, further hindering scalability and sustainability. Overcoming these challenges is critical to the future development of MHP-based technologies. Here, we present a novel water-based solvent system and synthetic approach for the preparation of size-controlled CsPbBr$_3$ perovskite nanocrystals in ambient air and at room temperature. The photoluminescence quantum yield (PLQY) of CsPbBr$_3$ perovskite nanocrystals (PNCs) exceeds 60%. To demonstrate the light to current conversion ability of our PNCs a series of photoconductors were prepared, with the best performing devices achieving a specific detectivity ($D^*$) of 1.2 x 10$^{11}$ Jones. Thus, this green, scalable, and low-cost approach offers a sustainable pathway for precise size and compositional control of MHP nanocrystals, opening new possibilities for environmentally friendly optoelectronic applications.

1. Introduction

Perovskite nanocrystals (PNCs) are promising candidates for next-generation light-emitting diodes (LEDs), photovoltaics (PVs), photodetectors (PDs), and other optoelectronic applications. Their exceptional properties—including high-purity, bright emission, tuneable absorption, ambipolar charge transport, long charge carrier diffusion lengths, and intrinsic defect tolerance—make them highly attractive for advanced device technologies[1]–[3]. The relative ease by which compositional control can be afforded through well-established synthesis methods permits facile doping, thus a broad range of bandgaps can be prepared whereby absorption and/or emission wavelength can be tuned to suit the application. Recent PNC developments have demonstrated tuneable emission across the visible spectrum and beyond *e.g.* CsSnI$_3$ with near-infrared emission[4] to CsPbCl$_3$ with blue-to-violet emission[5], and quantum yields (QYs) approaching 100%. Synthesis of PNCs broadly follows two well-reported synthetic methods, namely ligand-assisted reprecipitation (LARP)[6] and hot-injection (HI)[7]. LARP is generally considered to be more facile than HI, and more efficient, and as elevated temperatures are not required there is a greater potential for scale-up[8],[9]. The main limitations of the LARP method stem from the significant environmental and health risks posed by the solvents used, particularly the combination of dimethyl sulfoxide (DMSO) and dimethylformamide (DMF), which present challenges for scaling-up and mass production[10]. As PNCs continue to garner interest, there is a requirement to identify synthesis methods that retain the positive attributes of LARP and, in parallel, address the environmental and human health risks associated with the solvents employed. Despite this growing need there remain few reports focusing on PNC synthesis using green solvents or processes. Some notable examples include the work of Lu *et al.*, who developed natural deep eutectic solvents (NADES) to substitute DMF and ligands used in LARP, demonstrating the successful synthesis of $CsPbBr_3$ PNCs[11]. In 2019, Ambroz *et al.* demonstrated a room-temperature injection method for $FAPbBr_3$ and $CsPbBr_3$ PNCs using a phosphine-based ligand combination that enabled the dissolution of all precursors without the need for polar aprotic solvents such as DMF or DMSO[12]. Hoang *et al.* utilized an environmentally friendly ionic liquid, based on a series of methylammonium carboxylates, to prepare $MAPbBr_3$ PNCs[13] and Chatterjee *et al.* used menthol-based deep eutectic solvents (DESs) to prepare $CsPbX_3$ (X = I, Br, Cl) PNCs and nanoplates[14]. Water, the greenest of solvents has received some attention as a solvent for PNC synthesis, this despite the remarkably low solubility of lead halide compounds coupled with the widely held perception of the detrimental effects of water on metal-halide perovskites.[15],[16] In such reports lead halide solubilisation is achieved by the formation of water-soluble complexes, driven by the addition of large quantities of hydrohalic acids i.e. aqueous acids, however solubility still remains low reactions are carried out at the extremes of pH (often negative). Similarly, there are a handful of reports investigation post-deposition water treatments to tune surface properties. [17]–[20] Here, we report the development of a novel, water-based solvent capable of dissolving high concentrations, > 0.2 mol dm$^{-3}$ of otherwise insoluble lead halide compounds at room temperature and the subsequent development of a facile, rapid, efficient, and non-toxic synthetic route for the preparation of size-controlled $CsPbBr_3$ PNCs. Our solvent consists solely of water, β-alanine (βA, 3-aminopropanoic acid) and malic acid (MA, 2-hydroxybutanedioic acid). We have developed a primary solvent system that is environmentally sustainable, relying on naturally occurring components that are abundant, biodegradable and non-toxic thus posing minimal environmental risk and aligning with green chemistry principles.[21],[22]

Our solvent uniquely enables the aqueous dissolution of $PbBr_2$ through adduct formation, by interaction with the carboxylate end group of MA. The subsequent PNC synthesis is triggered by reaction with hydrobromic acid (HBr), following which the PNC surfaces are stabilised by the addition of common ligands and the prepared PNCs are dispersed in toluene ready for deposition. We rigorously investigate the role of MA, βA, PbBr$_2$ and HBr and their relative concentrations on PNC synthesis, resulting in the successful formation of size-controlled PNCs with sizes ranging from < 5 nm to 100 nm and photoluminescence quantum yields (PLQY) > 60 %.Therefore our novel, environmentally benign solvent provides a clean, scalable, and cost-effective strategy for precise control over the size of PNCs.

2. Results and Discussion

2.1. Developing an aqueous solvent system βA and MA are naturally occurring molecules, the former is synthesised in the human liver and is often consumed as a sports supplement whilst MA is the primary acid of many common fruits, the molecular structures of both are shown in Figure 1a. We first consider the molecular states that each molecule, and mixtures of the two, can adopt in aqueous solution, Figure S1. βA contains an amine and a carboxylic acid group that, in aqueous environments, can undergo proton transfer leading to the zwitterionic form containing ammonium and carboxylate charged end groups (pK$_a$ values of around 3.55 and 10.2 respectively)[23]. MA contains two carboxyl groups (pK$_{a1}$ of 3.4 and pK$_{a2}$ of 5.1)[24] which can be partially or fully deprotonated in water, forming carboxylate groups.

To determine the molecular states present in each solution we used attenuated total reflectance-Fourier transform infrared spectroscopy (ATR-FTIR) initially on solid samples of βA and MA, Figure S2. In βA the absorption cented at 1631 cm$^{-1}$ and the shoulder at 1650 cm$^{-1}$ can be assigned to $_{vas}$COO$^-$ with and $_{δas}$NH$_3^+$ respectively ($_{δas}$NH$_3^+$ also at 1567 cm$^{-1}$ and $_{δs}$NH$_3^+$ also at 1504 cm$^{-1}$) indicating βA is in a zwitterionic form[25]. For MA, the absorption at 1680 cm$^{-1}$ and the ajacent absorptions at 1714 and 1737 cm$^{-1}$ collectively correspond to $_v$(C=O) of dimeric COOH formed between MA molecules[26]. Aqueous solutions of βA and MA were prepared each with mole fractions ($X_{βA}$ /$X_{MA}$) of 0.091, resulting in solutions with pH values of 7.48 and 1.32 respectively, Table S1. In solution, significant changes in the FTIR spectra of both molecules are seen Figure S2. Considering the FTIR spectra of the solution for βA we anticipate, owing to the zwiterionic nature, that the molecule will be heavily solvated by water. The discrete absorptions observed around 1504 – 1650 cm$^{-1}$ are replaced by a broad absorption band centred at 1560 cm$^{-1}$ that indicates a strong interaction of water with these charged end groups. In solution, MA shows a similar broad band centred around 1714 cm$^{-1}$ that shows the dimers formed in the solid are disrupted and the carboxlate groups are solvated. Solutions of both molecules were then prepared where the concentration of each was systematically varied, Table S1, resulting in changes in the FTIR spectra, Figure S3, that can be attributed to solution composition and the relative concentrations of each molecule.

As conductivity is derived from the concentration of dissolved ions in solution *i.e.* the non-neutral molecular states, we measured the pH of our solutions, **Figure 1**b, to further interrogate the molecular states of βMA. For βA the zwitterionic nature means in solution the molecule will be electrically neutral *i.e.* a form that will not contribute to conductivity[27] and this is supported by our results (0.1 mS cm$^{-1}$). We then mix solutions of the two molecules, initially increasing the quantity of MA added to a fixed quantity of βA, Table S1. The addition of MA ($X_{MA}$ = 0.015) results in a sharp increase in conductivity (15.5 mS cm$^{-1}$) driven by the formation of carboxylate groups caused by the interaction of MA with βA. Systematically raising the amount of MA to $X_{MA}$ = 0.043 results in increased conductivity, (18.2 mS cm$^{-1}$), however further increases in the quantity of MA, that is until equivalent mole fractions of MA and βA are reached results in a subtle fall in conductivity, Table S1. In contrast, pure solutions of MA have a measured conductivity of 5.7 ± 1.2 mS cm$^{-1}$ and conductivity steadily increases as βA is added. We note that when the relative quantities of MA < βA the solution conductivity is greater than when MA > βA and whilst the solution only containing MA has the lowest pH the conductivity is not derived alone from pH, rather the molecular forms of βA and MA in solution.

2.2. Assessing the solubility of perovskite precursors in our aqueous solvents

We now consider the solubility of CsBr and PbBr$_2$ in our solvent mixtures noting that PbBr$_2$ is insoluble in water. We observe that solubility is only achieved in three of the compositions studied, shown in Figure S4 and detailed in Table 1. At room temperature, solutions containing equimolar quantities of CsBr and PbBr$_2$ **at concentrations up to** 0.3 mol dm$^{-3}$ can be solubilised, however all results herein reported contain equimolar quantities at a concentration of 0.2 mol dm$^{-3}$ *i.e.* away from the solubility limit. Unless otherwise stated all data reported were obtained from the aqueous Solvent 1**, composition detailed** in Table 1. We propose that the insoluble nature of PbBr$_2$ is overcome through adduct formation between the carboxylate groups of MA and PbBr$_2$, Figure S5, whereby these negatively charged groups function as a Lewis base enabling dissolution, Figure 1c.

Table 1. Summary of the aqueous solvent formulations in which both CsBr and $PbBr_2$ are solubilised. Comprehensive details of all solvent compositions investigated are provided in Table S1.

| Solvent | MA mole fraction ($X_{MA}$) | βA mole fraction ($X_{βA}$) | $H_2O$ mole fraction ($X_{H2O}$) | pH | Conductivity (mS cm$^{-1}$) |
|---|---|---|---|---|---|
| 1 | 0.015 | 0.090 | 0.895 | 4.46 ± 0.1 | 15.5 ± 0.56 |
| 2 | 0.029 | 0.088 | 0.883 | 4.18 ± 0.1 | 18.2 ± 1.05 |
| 3 | 0.043 | 0.087 | 0.870 | 3.91 ± 0.1 | 18.2 ± 1.15 |

We investigate this by drying a mixture of CsBr, $PbBr_2$ and our solvent and conducting FTIR, **Figure 1**d. The 1650 – 1000 cm$^{-1}$ region of the dried solid closely resembles that of pure βA, the bands at 1633 cm$^{-1}$ and 1573 cm$^{-1}$ ($_{vs}COO^-$ with and $_{δas}NH_3^+$ respectively) are unchanged, indicating minimal interaction between βA and the precursors whilst $_v$(C=O) of MA shifts considerably, indicating there is an interaction between this group and the precursors. To confirm that adduct formation is between the deprotonated carboxyl group of MA and $PbBr_2$ we attempted to dissolve $PbBr_2$ in aqueous solutions of pure MA and pure βA, however neither solution could dissolve $PbBr_2$ (Figure S6). The addition of small quantities of base (NaOH or $NH_4OH$) to these solutions did result in dissolution in the case of MA but not βA, Figure S6. $^1$H NMR offers further evidence of an interaction of MA with $PbBr_2$ **(Figure S7)**. Compared to MA alone, mixing with βA resulted in an upfield shift of the proton on the carbon bearing the hydroxyl group, indicative of carboxylate deprotonation. The addition of $PbBr_2$ resulted in a downfield shift, suggesting interaction of carboxylate with Pb. This dissolution mechanism is also applicable in acids with similar structure as MA *e.g.* succinic acid (SA) Figure S8.

Having developed an understanding of this unusual solubility we turn to consider the nature of the species formed in solution. Using dynamic light scattering (DLS) we probed our neat solvent system and our solvent with the addition of CsBr and $PbBr_2$. Although MA and βA exhibit high solubility in $H_2O$ we predicted that in our solvent systems, owing to the interaction of the molecules at relatively high concentration and their individual interactions with water, that they may possess some colloidal properties. DLS analysis shows the dynamic size of our neat solvent to be around 1 nm. Following the addition of CsBr and $PbBr_2$ a portion of the solvent remains, presumably not involved in adduct formation, however there is a large population of species with a mean size around 600 nm that we attribute to the adduct formed with $PbBr_2$, **Figure 1**e.

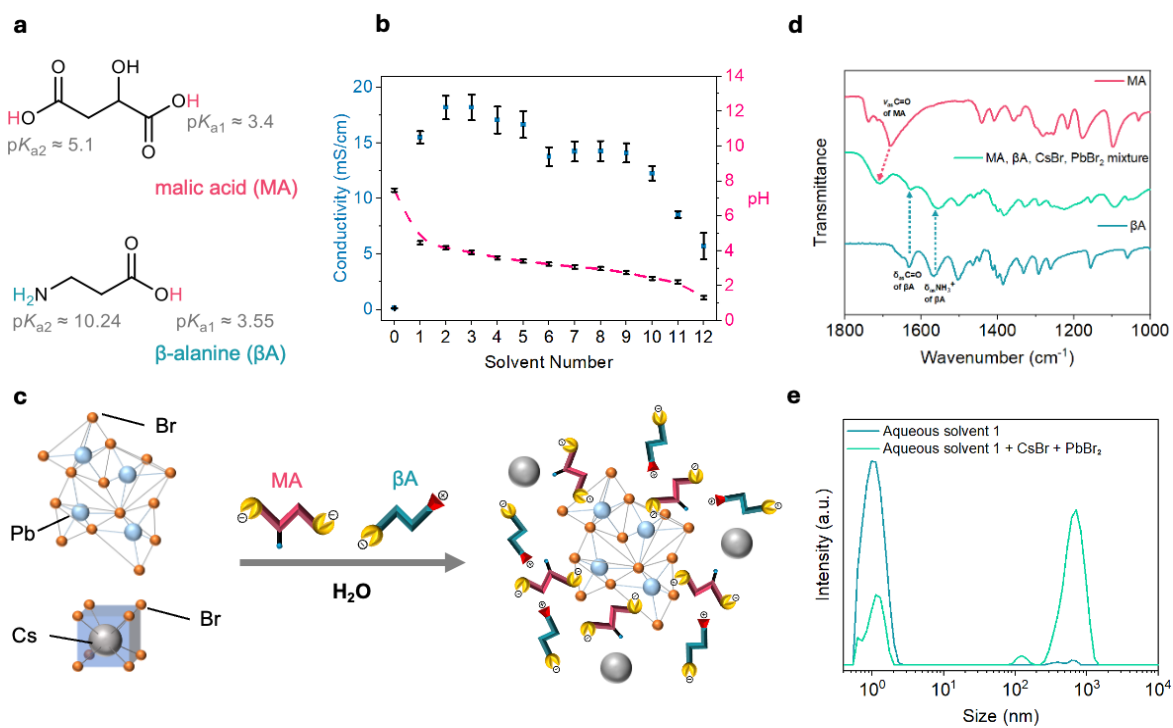

**Figure 1.** a) Molecular structures and p$K_a$ values of malic acid (MA) and β-alanine (βA). b) Conductivity and pH measurements of aqueous solutions containing MA, βA, and their mixtures (see Table S1). c) Schematic illustration of the proposed dissolution mechanism for CsBr and PbBr$_2$ in the MA/βA solvent system. **d)** Fourier transform infrared spectra of βA, MA, and the dried product obtained after mixing the solvent with CsBr and PbBr$_2$, indicating coordination interactions. e) Dynamic light scattering data showing changes in particle size distribution before and after precursor dissolution.

## 2.3. Understanding Growth and Nucleation Processes

To synthesize PNCs, it is necessary to precipitate the perovskite from the precursor solution by disrupting the interactions between MA and PbBr$_2$. Based on our model of precursor dissolution we propose a mechanism that utilises protons (HBr) to disrupt the adduct formed between MA and PbBr$_2$, driven by the conversion of carboxylate groups into carboxylic acid groups, Figure S9, followed by precipitation of the desired perovskite. Owing to the high solubility of CsBr in water the addition of a proton source to the mixture will likely only precipitate PbBr$_2$ and not the desired perovskite. Therefore, it is necessary to introduce a co-solvent with the proton source that has good miscibility with water but poor solubility with our precursors. We identified and investigated three such solvents, methanol, isopropanol (IPA), and acetone -however acetone was the only solvent that successfully facilitated the formation of PNCs (**Figure S10**).

The synthesis proceeds by mixing known quantities of concentrated HBr with 3mL of acetone and injecting 100 μL of our aqueous precursor mixture to this solution. Our precursor contains deprotonated MA and βA both insoluble in acetone, their mixing results in the formation of gel-like aggregates and phase separation, Figure S11. However, protonated MA and βA are soluble in acetone, hence mixing with HBr breaks the adduct formed between MA and $PbBr_2$, thereby releasing $PbBr_2$ to react with $Cs^+$ and $Br^-$ ions forming insoluble PNCs. The proposed reaction pathway is outlined in **Figure 2**a. The synthesised PNCs are unstable in polar solvents, acetone and water in this case; therefore, surface passivation is required to achieve stability. Passivation is achieved by adding the synthesised PNC solution to a mixture of oleylamine (OAm) in toluene, followed by centrifugation. The supernatant is removed, and the precipitate redispersed in a mixture of OAm, oleic acid (OA) and toluene creating a stable dispersion. Ligand addition is critical, without which aggregation occurs, leading to challenges in dispersion and the loss of quantum confinement effects (Figure S12).

To better understand PNC growth, we studied the role HBr plays in the synthesis. A series of HBr-acetone solutions were prepared in which the HBr concentration was systematically varied from 1.5 to 4.4 % v/v, **Table 2**. Distinct differences in reaction rate were observed as HBr concentration was changed as indicated by the formation of orange-coloured solution. At the lowest HBr concentration the reaction proceeds over 2-3 seconds and gets faster until 3.7 % v/v is used, at which point the colour change is instantaneous upon mixing. Further increasing the concentration results in the precipitation of bulk $CsPbBr_3$ (Figure S13). Transmission electron microscopy (TEM) was carried out on the PNCs prepared over the HBr concentration range, **Figures 2**b. Analysis of the images reveals a small increase in size with increased HBr concentration accompanied by a slight reduction in dispersity **Table 2**. We propose that size increase is driven by the rate of the reaction of MA with HBr, resulting in $PbBr_2$ dissociating from the adduct and forming PNC precipitates. At higher HBr concentrations, $PbBr_2$ is released to the reaction quickly, thus the nucleation concentration ($C_{cr}$) is achieved and exceeded rapidly ($t_{H1} < t_{L1}$). Nucleation and subsequent nanocrystal growth then consumes the released $PbBr_2$ and thus the $PbBr_2$ concentration quickly falls back to below $C_{cr}$ ($t_{H2} < t_{L2}$). In this regime there is no further nucleation, allowing more time for crystal growth resulting in larger PNCs, (**Figure 2**c). The changes in size and dispersity were also reflected in the photoluminescence spectra (PL), Figure S14. As the PNC size increases, a redshift in the spectra occurs with an accompanying narrowing of the emission, **Table 2**. In some samples there is some signal observed around 450 nm that is likely arising from broad band defect states owing to halide vacancies[28]. The photoluminescence quantum yields (PLQYs) fall by almost an order of magnitude as PNC size/HBr concentration increased, **Figure 2**d, with PLQY falling from 25.3% (1.5 % v/v) to 3.3% (3.7% v/v).

**Table 2.** The average size, size dispersity, PL emission maxima ($\lambda_{max}$) and full-width at half-maximum values of PNCs as a function of HBr concentrations and solvent ageing.

| HBr% v/v | *Fresh HBr-acetone* | | | | *1-hour aged HBr-acetone* | | | |
|---|---|---|---|---|---|---|---|---|
| | Size (nm) | Dispersity (nm) | $\lambda_{max}$ (nm) | FWHM (nm) | Size (nm) | Dispersity (nm) | $\lambda_{max}$ (nm) | FWHM (nm) |
| 1.5 | 30.9 | ± 30.1 | 516.1 | 21.5 | 6.3 | ± 1.2 | 498 | 31.0 |
| 2.3 | 34.0 | ± 15.9 | 518.9 | 20.8 | 9.9 | ± 4.1 | 513 | 24.0 |
| 3.0 | 39.4 | ± 26.9 | 517.0 | 21.2 | 10.1 | ± 4.9 | 512 | 24.0 |
| 3.7 | 38.1 | ± 25.7 | 518.9 | 20.4 | 45.2 | ± 22.0 | 515 | 23.5 |
| 4.4 | *N/A* | *N/A* | *N/A* | *N/A* | *N/A* | *N/A* | *N/A* | *N/A* |

To explore the recombination processes affecting PLQY, we carried out time-correlated single photon counting (TCSPC) experiments. The measured data were fitted to biexponential decays **Figure 2**e (see Table S2 for fitting parameters) where the first term reflects a fast decay phase, assigned to trapping into non-radiative deep traps, and a second decay phase assigned to bimolecular recombination of long-lived free carriers, Equation S1. The short-lived lifetime ($\tau_1$) is related to the recombination of excitons initially generated upon photon absorption whilst the long-lived carrier lifetime ($\tau_2$) is related to exciton recombination at surface states[29]. PNCs prepared from 1.5 %v/v HBr showed the highest PLQY, where the short- ($\tau_1$) and long-lived carrier lifetimes ($\tau_2$) were 5.4 ns and 37.3 ns respectively. PNCs prepared by precipitating using 2.3 and 3.0% 5v/v HBr had similar PLQY values to PNCs prepared with 1.5% v/v HBr, and comparable $\tau_1$ lifetimes, however their $\tau_2$ values were considerably greater. From these data we have calculated the radiative ($k_r$) and non-radiative ($k_{nr}$) recombination rate constants, Table S2. Generally, carrier lifetime depends not only on the intrinsic properties of the perovskite but additionally on the defect density, composition and surface chemistry[30],[31]. The observed reduction in $k_{nr}$ and $k_r$ seen as HBr concentration increases is attributed to the accompanying increase in PNC size, where larger PNCs become affected by surface defects[32],[33].

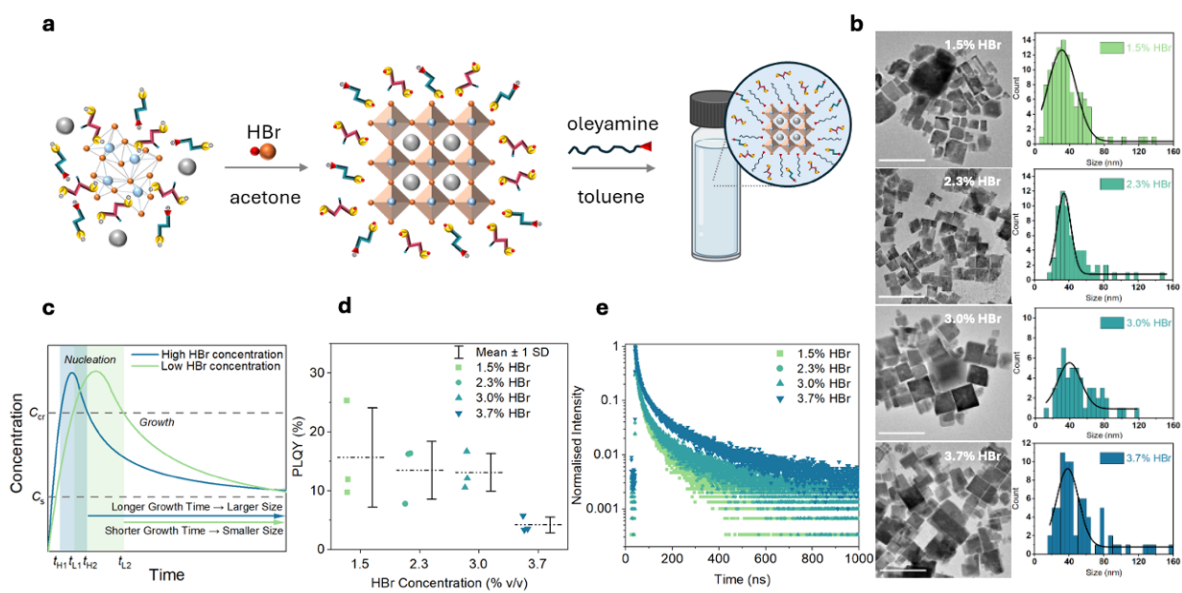

**Figure 2.** a) Schematic of the proposed reaction pathway: solubilisation of $PbBr_2$ via adduct formation in water, followed by precipitation and ligand-mediated surface passivation. All subsequent data correspond to PNCs precipitated using 1.5%, 2.3%, 3.0%, and 3.7% v/v HBr in acetone. b) Transmission electron microscopy images of PNCs (scale = 50 nm) and corresponding statistical size analysis. c) LaMer diagram illustrating the effect of HBr concentration on nucleation and growth dynamics. d) Photoluminescence quantum yield of PNCs as a function of HBr content. e) time-correlated single photon counting measurements showing exciton lifetime dynamics under varying synthesis conditions.

2.4. Microstructural and compositional characterisation

The X-ray diffraction (XRD) patterns of PNCs prepared with varying HBr concentrations are shown in Figure 3a. At 1.5% v/v HBr the (110), (112), (004), and (220) peaks were observed, consistent with the stabilisation of orthorhombic $CsPbBr_3$ (ICDD 01-090-0544). Additionally, peaks from the (113), (300), (024) and (214) planes were observed, which can be attributed to hexagonal $Cs_4PbBr_6$ (ICDD 04-014-8071). As the concentration of HBr is increased, the quantity of the hexagonal phase decreases until the formation of phase pure $CsPbBr_3$ at 3.7 % v/v HBr is observed, Figure 3b.

To study the surfaces of our PNCs, FTIR and X-ray photoelectron spectroscopy (XPS) were used in combination. The FTIR spectra of OA, OAm, βA, MA, and prepared PNCs are shown in **Figure 3**c. In the spectra of OA and OAm, two bands from 2800 to 2990 cm$^{-1}$ represent $CH_2$ stretching[34]. These bands were also present in the spectrum of PNCs, indicating the presence of OA and OAm on the surface. Additionally, the presence of a broad band from 2800 to 3300 cm$^{-1}$ is characteristic of H-bonding, absent in OAm and OA but present in MA and βA. The band at 1567 cm$^{−1}$ in the βA spectrum is attributed to NH$_3^+$ stretching, while the band at 1097 cm$^{−1}$ corresponds to C-O stretching mode in MA [26]. These bands were also observed in the spectrum of PNCs, although they are less prominent, suggesting residual amounts of MA and βA still exist on the surface of the PNCs.

High-resolution XPS spectra obtained from PNCs prepared using 1.5% v/v HBr are shown in **Figure 3**d-e. The N 1s core level was fitted with two components at 399.9 eV and 401.6 eV, representing −NH$_2$ and −NH$_3^+$ respectively[35]. In traditional LARP methods, where OA and OAm have been used as surface ligands, both -NH$_2$ and -NH$_3^+$ are reported in XPS results, originating from OAm. In our case the reported intensity of the peak at 399.9 eV was higher than 401.8 eV, which was attributed to a greater concentration of surface -NH$_2$ [34]. In our data the intensity of -NH$_3^+$ is greater than -NH$_2$, indicating -NH$_3^+$ from βA also exists on the surface, consistent with FTIR results. In the O 1s spectrum, **Figure 3**e, the broad peak from 534.5 eV to 529 eV is fitted with four components. A small peak at 529.7 eV is likely originating from PbO, is consistent with the Pb 4f spectrum (Figure S17), although the quantity present is small. The peaks at 531 eV and 533 eV are attributed C=O and -OH respectively, which is present in both OA and MA. The peak at 532 eV arises from the hydroxyl group of MA *i.e.* indicating the presence of residual MA in the PNC, again consistent with the ATR data, **Figure 3**c. The Pb 4f doublet is shown in Figure S17, which also indicate the presence of small quantities of metallic Pb, indicated by the additional peak at 136.0 eV. The main, broad peak centred at 137.7 eV can be fitted to Pb$^{2+}$ (from the perovskite)[36]. High resolution transmission electron microscopy (HRTEM) was carried out on 1.5% v/v HBr sample. Figure S18. The presence of well-defined lattice fringes in the PNCs confirm the crystalline nature of these samples, from the corresponding data the lattice spacing was calculated to be 0.41nm, consistent with the (002) plane [37].

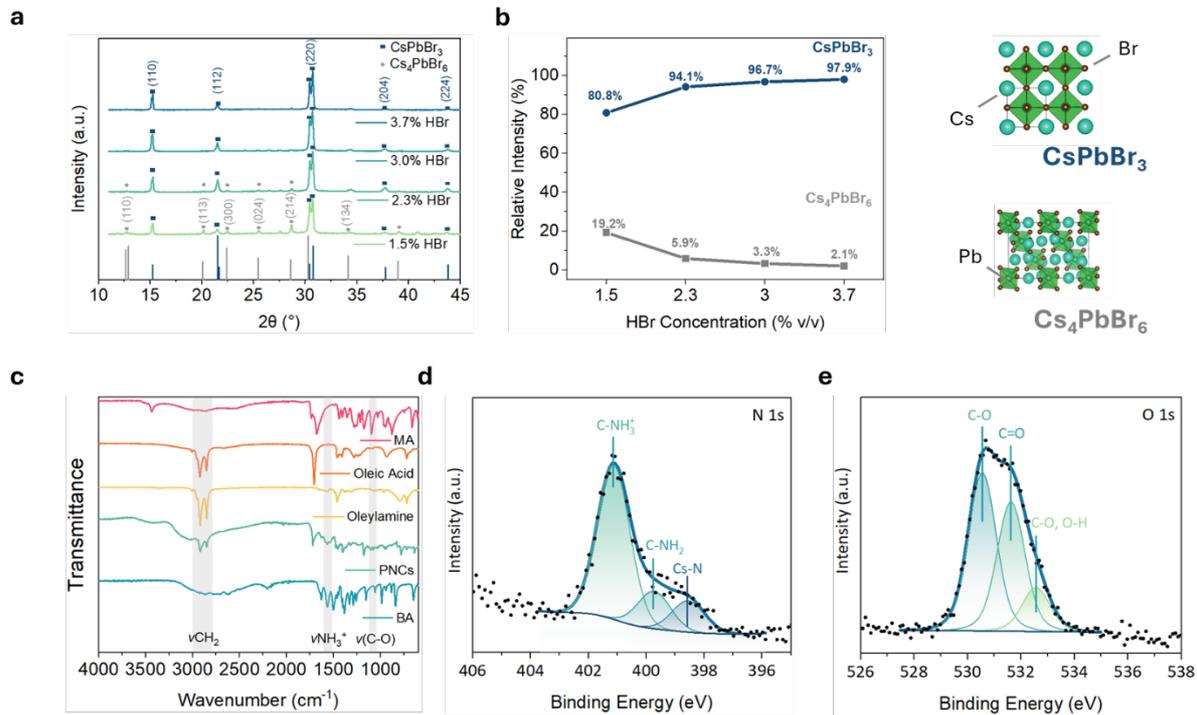

**Figure 3.** a) X-ray diffraction patterns of PNCs as a function of HBr concentration, alongside reference patterns for CsPbBr$_3$ (ICDD 01-090-0544, dark blue) and Cs$_4$PbBr$_6$ (ICDD 04-014-8071, grey). b) Estimated phase composition of PNCs, determined from quantitative analysis of the XRD data. c) Fourier transform infrared spectra of oleylamine, oleic acid, β-alanine (βA), methylammonium (MA), and CsPbBr$_3$ PNCs. and fitted high-resolution X-ray photoelectron spectroscopy spectra for d) N 1$s$ and e) O 1$s$.

## 2.5. Ageing effects of precipitating solution mixture

We observed that if the HBr-acetone solution was not used immediately after its preparation that it developed a brown colour, ascribed to the acid-catalysed enolisation of acetone, if this was then used to prepare PNCs their properties changed considerably. To elucidate the origins of this we prepared solutions of HBr and deuterated acetone (acetone-d$_6$) and recorded $^1$H-NMR spectra over time. The $^1$H NMR spectra (1.5% v/v HBr) of the fresh 0-, 1-, 5- and 96-hour aged mixtures are shown in **Figure 4**a. The quintet peak at 2.05 ppm originates from the presence of small amounts of residual protonated species in the acetone-d$_6$ and is used as a reference. Since there was also trace water in the mixture, the proton exchange between HBr, H$_3$O$^+$ Br$^-$, and water molecules was fast, resulting in a composite singlet peak located at 8.27 ppm at 0-h aging[38]. The chemical shift of this singlet peak is given by **Equation 1**, where $f$ and $\delta$ are the fraction and chemical shift of the corresponding component, respectively:

$$\delta_{av}(H) = f_{H_2O}\delta_{H_2O} + f_{H_3O^+}\delta_{H_3O^+} + f_{HBr}\delta_{HBr} \tag{1}$$

The chemical shift of this peak is therefore related to the acidity of the precipitating solution. $^1$H NMR was also carried out on 1.5 and 2.3 % v/v HBr solutions immediately after mixing (0-hour), Figure S19. The composite peak of 2.3% v/v HBr showed a larger chemical shift than that of 1.5% HBr v/v equivalent *i.e.* at higher $H_3O^+$ concentration the composite peak shifted downfield. **Figure 4**a shows that with increasing time the composite peak shifts upfield, from 8.27 ppm (0-hours) to 6.7 ppm (48-hours).

The UV-vis absorption spectra of the 0-hour and aged HBr-acetone mixtures are displayed in **Figure 4**c. For the 0-hour samples the absorption onset was located at 260 nm, consistent with pure acetone in water. However, with increased time the absorption onset was shifted to 250 nm which was attributed to the formation of 1-propen-2-ol ($CH_2C(OH)CH_3$) following the enolisation of acetone, consistent with $^1$H NMR observations. The TEM images of PNCs prepared using HBr aged for 1-hour are presented in **Figure 4**b. For PNCs made from 1-hour aged 1.5% v/v HBr, a regular cubic morphology was observed, and the average size calculated to be 7.2 nm, which is smaller compared to when a fresh acetone-HBr solution was used, and like the case when unaged HBr-acetone was used, there is a systematic increase in PNC size with increasing HBr concentration, **Table 2**, However, aging the precipitating solution results in a reduction in the absolute dimensions when compared with unaged equivalents. The PL data, **Figure 4**d, reflect the observations made by TEM. For the smallest PNCs the PL emission maximum was located at 498 nm and the narrow size distribution resulted in relatively narrow FWHM (23.3 nm). A progressive redshift in emission was observed as the size increased, with maxima up to 515 nm occurring for the largest PNCs, **Table 2**. According to the $^1$H NMR data (**Figure 4**a) the composite peak of $H_3O^+$, $H_2O$ and HBr shifted upfield with increased ageing time, indicating that the fraction of $H_3O^+$ decreased with time. This could be caused by $Br^-$ being consumed by oxidation to $Br_2$, which would be accompanied by the presence of a brown colour as observed (see Figure S20). Another possibility could be that the concentration of molecular HBr was increased and correspondingly the fraction of $H_3O^+/Br^+$ reduced. In both cases the reduction in concentration of $H_3O^+$ results in a reduction of the rate of $PbBr_2$ released into the solution *i.e.* nucleation is slowed, thus PNC size is reduced when the aged HBr solutions are used.

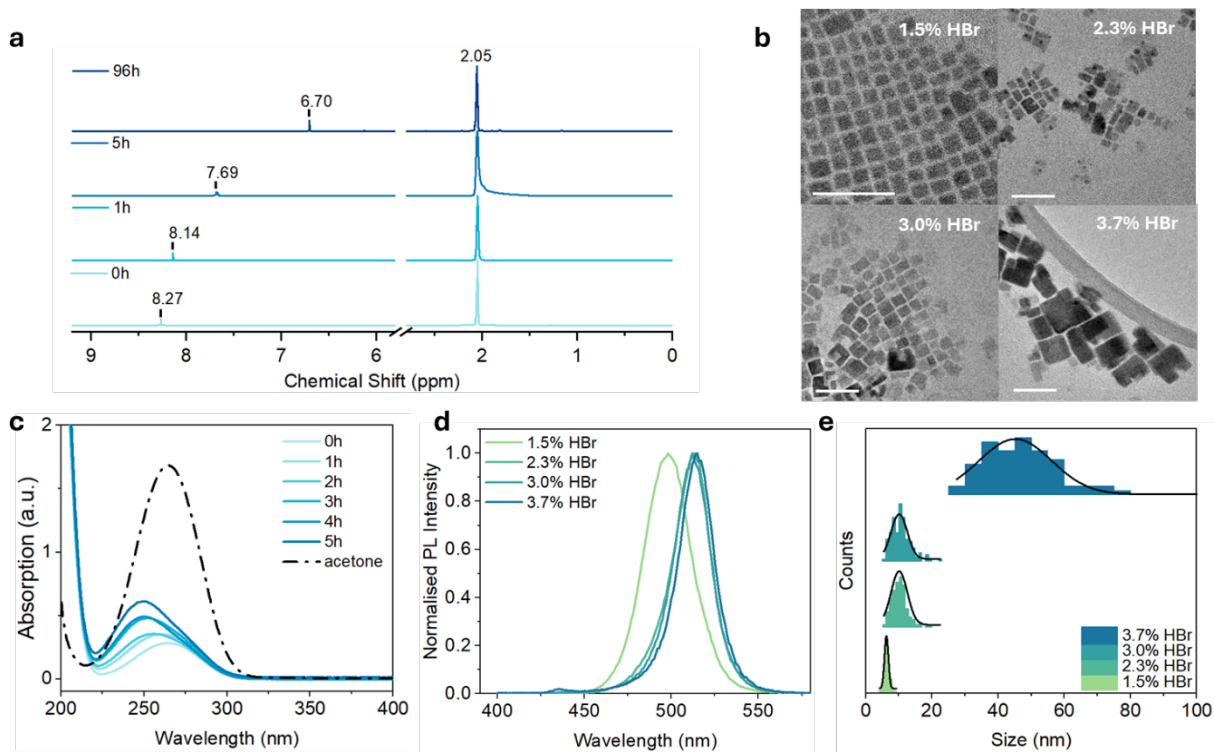

**Figure 4.** a) Time-resolved $^1$H NMR spectra of precipitating solutions containing 1.5% v/v HBr, showing chemical evolution upon ageing. b) TEM images of perovskite nanocrystals (PNCs) synthesised from solutions aged for 1-hour with varying HBr content (1.5 - 3.7% v/v), scale bar=50 nm. c) UV–vis absorption spectra of 1.5% v/v HBr precipitating solutions aged over a 5-hour period. d) Photoluminescence spectra of PNCs prepared from 1-hour aged solutions containing 1.5 - 3.7% v/v HBr. e) Size dispersity measurements of PNCs derived from the same ageing conditions, highlighting the sensitivity of nanocrystal uniformity to precursor solution composition.

Having obtained an understanding of the role of the precipitating solution we consider the impact of changing the quantities of βA and MA in our aqueous solvent, noting that the solubility of PbBr$_2$ and CsBr could be achieved in three of our solvent compositions (**Table 1**) The results presented thus far have been obtained only using Solvent 1. We turn to the use of Solvents 2 and 3 in our synthesis Using precipitating solutions containing 1.5-3.7 % v/v HBr aged for 1-hour, the representative TEM images are shown in **Figures 5**a-b. It is apparent that the samples follow the same trend seen previously, increasing in size with increasing HBr concentration (**Figures 5**a-d). There are little appreciable differences in the size of PNCs prepared in each precipitating solution however the HBr concentration is having a significant impact. In each solvent the mean diameter is around 7-8 nm when 1.5 % v/v HBr is used for precipitation increasing to 27-28 nm with 3.7 % v/v HBr, although notably the dispersity of the PNCs prepared using Solvent 3 is greater in every case. The anticipated red-shift in PL emission is seen in both systems as PNC size increases, **Figure 5**e, comparable with previous results, **Figure 4**d.

We now consider the XRD data from these PNCs prepared using only the 1-hour aged 1.5 % v/v HBr, **Figure 5**f. For all three solvents several phases are identified, primarily $CsPbBr_3$ (ICD 01-090-2544), $CsPb_2Br_5$ (ICDD 01-090-0382) and $Cs_4PbBr_6$ (ICDD 04-014-8071) with no discernible differences observed with varying solvent, however when the HBr concentration is increased, the quantities of $Cs_4PbBr_6$ and $CsPb_2Br_5$ are reduced, Figure S21, to the extent that when 3.7% v/v HBr is used the PNC composition is predominantly $CsPbBr_3$, consistent with the observations made when increasing HBr concentration in unaged precipitating solution, **Figure 3**a.

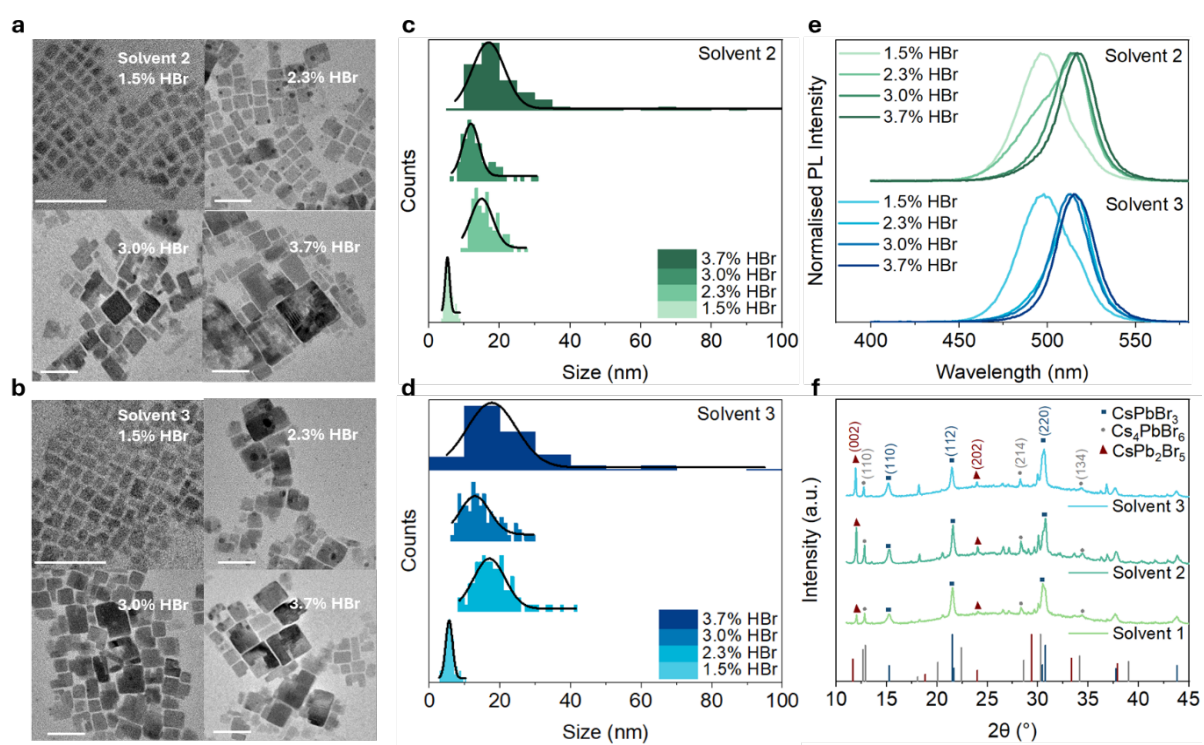

**Figure 5.** All data shown were obtained using 1-hour aged HBr–acetone solutions containing 1.5–3.7% v/v HBr, unless otherwise noted. a-b) Transmission electron microscopy images of PNCs synthesised using Solvent 2 and Solvent 3, respectively (scale = 50 nm). c-d) Size distribution histograms corresponding to PNCs in panels a and b, derived from particle analysis of TEM images. e) Photoluminescence spectra of PNCs synthesised using Solvent 2 and Solvent 3. f) X-ray diffraction patterns of PNCs prepared using Solvents 1–3, compared to reference patterns from ICDD: $CsPbBr_3$ (01-090-2544, blue), $Cs_4PbBr_6$ (04-014-8071, grey), and CsBr (01-090-0382, red).

The PLQY data for PNCs prepared using Solvent 1 and precipitated from solutions containing 1.5 - 3.7 % v/v HBr aged for 1-hour are shown in Figure S22. A reduction in PLQY was observed as the HBr concentration increased, consistent with previous observations however the PLQY values for all samples prepared using 1-hour aged HBr solutions exceed those prepared using fresh equivalents. Additionally, the maximum PLQY values showed a significant enhancement for the 1-hour mixture compared with the fresh solution, with averages of 56% and 16% respectively. Figure S22 and **Figure 2**d. To gain insight into the PLQY enhancement TCSPC was again employed where the PLQY data were fitted to biexponential decays Figure S23 (see Table S4 for fitting parameters). When compared to data obtained from PNC prepared using fresh HBr-acetone solutions there is a small reduction in the $\tau_1$ component, from 5 ns to 4 ns, and a significant reduction in the slow component ($\tau_2$), falling from 40 ns to 20 ns. This indicates that the carriers show higher tendency to recombination in PNCs prepared from the aged HBr-acetone solutions. This is consistent with the values of the recombination rates calculated from Equations S2-3. Both radiative and non-radiative recombination rates increased for PNCs prepared with aged HBr-acetone solvents, possibly associated with a stronger quantum confinement effect induced by the smaller size of the PNCs (~7 nm), forcing excitons generated after photoexcitation to radiatively recombine at a faster rate[39], [40], explaining the higher PLQY. We have described the limitations on scale-up imposed by conventional PNC synthesis methods and have implicitly highlighted our aqueous solvent system as being intrinsically more amenable to scale-up owing to the non-toxic nature of the components, the sustainability of the materials, and potential for simpler waste management compliance. To demonstrate the potential of our novel system for scale-up we have adapted the synthesis conditions to increase the scale by a factor of 30 relative to that reported in the rest of our manuscript, Figure S24. Recognising the importance of achieving tuneable optoelectronic properties, we have conducted post-synthetic anion exchange reactions. These experiments successfully demonstrate bandgap tunability, seen through systematic shifts in the emission spectra and, the predicted shifts in the XRD patterns as anion substitutions are carried out (Figure S25).

## 2.6. Photoconductor device measurements

Photoconductors were prepared to probe the light-to-current conversion capability of the PNCs. The device structure is illustrated in the inset in **Figure 6**a. The PNCs used in this study were synthesised from 1-hour aged 1.5% v/v HBr (Solvent 1), and measurements were conducted in air and at room temperature. The devices were tested under illumination from a 530 nm LED, with the power intensity ($P_{in}$) varied from 0.37 to 21.4 mW cm$^{-2}$. The current densities of the devices were measured with voltage range between -20 and +20 V and revealed a clear, monotonic increase in measured current density with increasing illumination intensity, **Figure 6**a. The responsivity was found to increase with increasing bias, **Figure 6**b. The maximum responsivity of 0.019 A W$^{-1}$ was achieved at -20 V under 21.4 mW cm$^{-2}$ illumination. The specific detectivity ($D^*$) of the photoconductors was evaluated by measuring the noise level with an applied bias between -1 and -20 V[41]. The noise spectral density as a function of frequency is shown in **Figure 6**c. The calculation of $D^*$ was performed by utilising the noise spectral density at 10 Hz, which resulted in a $D^*$ value of 1.2 x 10$^{11}$ Jones (at -20 V and 21 mW cm$^{-2}$ illumination), **Figure 6**d. The results obtained for our aqueous-processed PNCs are consistent with existing literature reports using common, yet toxic solvents[42]–[44], further emphasising the potential of preparing optoelectronic devices these benign solvents.

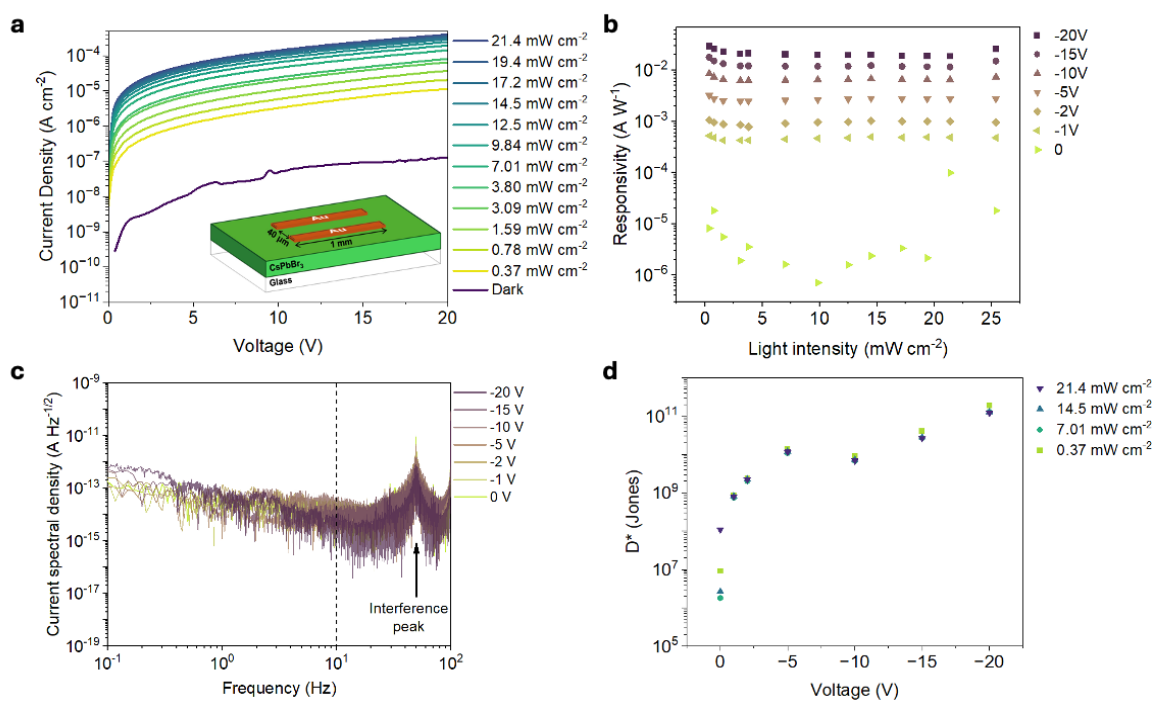

**Figure 6.** a) Current density as a function of incident light intensity, measured under different applied biases. Inset shows schematic illustration of the device architecture. b) Calculated responsivity of the devices as a function of light intensity and applied bias. c) Noise spectral density as a function of applied bias. d) Specific detectivity ($D^*$) calculated from responsivity and noise data, as a function of light intensity and bias.

## 3. Conclusion

We report the development of a novel aqueous, environmentally benign solvent system and synthetic protocol for the preparation of size- and composition-controlled metal halide PNCs. The solvent comprises only water, MA, and βA - all naturally occurring and non-toxic species that pose no risks to human health or the environment. Central to the process is the formation of a soluble adduct between MA and $PbBr_2$, enabling dissolution of an otherwise water-insoluble precursor. Subsequent mixing with HBr in acetone disrupts the adduct, driving the rapid precipitation of PNCs. Detailed spectroscopic analysis using FTIR and $^1$H NMR reveals specific interactions between solvent components and precursor species, allowing us to propose a mechanistic framework for crystal growth and identify the parameters governing PNC size and dispersity. We also uncover a solvent ageing effect within the HBr-acetone phase that significantly influences nanocrystal shape, size, and uniformity. The resulting PNCs exhibit strong photoluminescence, with quantum yields exceeding 60% in selected samples. Imaging by TEM confirms size tunability under different synthetic conditions, with size dispersity reflected in the breadth of the PL spectra. To demonstrate optoelectronic functionality, we fabricated proof-of-concept photodetectors, which display pronounced responses to both light intensity and applied bias, achieving a maximum specific detectivity of $1.2 \times 10^{11}$ Jones. Notably, the entire synthesis is performed in ambient air, in water, and without the need for any thermal treatment. This establishes a simple and sustainable route for the preparation of PNCs, with clear potential for scalable and sustainable manufacturing.

## 4. Experimental Section

*Materials*: Malic acid (DL-MA, 98%), β-alanine (βA, 99%), caesium bromide (CsBr, 99.9%,), lead bromide ($PbBr_2$, 99.999%), acetone (ACS reagent, 99.5%), toluene (ACS reagent, 99.5%) and methyl acetate (MeOAc, 99%) were purchased from Sigma Aldrich. Hydrobromic acid (HBr, 48% in water) was purchased from fluorochem. Oleylamine (OAm, > 50%) and oleic acid (OA, > 85%) were purchased from TCI.

*Preparation of solvent system*: Aqueous solutions of MA, βA were prepared at room temperature over the concentration range detailed in Table S1. $CsPbBr_3$ precursor solutions were prepared by dissolving equimolar quantities of CsBr and $PbBr_2$ into solutions of mixture of MA, βA and water and stirred at room temperature for 30 minutes to ensure complete dissolution. The solvent remains stable in ambient environments (fume hood) for durations exceeding six months.

***Synthesis and purification of CsPbBr₃ nanocrystals***: The entire synthesis process was carried out under ambient conditions and at room temperature. 3 mL acetone was transferred to a small vial and to it added different volumes of hydrobromic acid (48% v/v). Either 100 μL or 75μL of aqueous solvent (Solvents 1-3) containing MA and βA was added into the HBr-acetone mixture followed by shaking to precipitate the PNCs. The resultant solutions were mixed with 2mL toluene containing 50 μL OAm and centrifuged at 5000 rpm for 3 minutes. The supernatant was discarded, and a further 2 mL toluene containing 50 μL OAm and 100 μL oleic acid (OA, 99%, TCI) were added to the precipitate, resulting in a stable dispersion. A series of further toluene washing-centrifuge cycles were carried out (3000-5000rpm, 3-5 min) to reduce PNC dispersity resulting in a stable dispersion of PNCs in toluene.

***Modifying precipitating solvent volume***: As HBr concentration was having a direct influence on PNC properties, we investigated the impact of reducing the volume precursor solution from 100 μL to 75 μL whilst varying HBr concentration (Table S3), with representative TEM images are shown in Figure S15 and the PL spectra in Figure S16. At HBr concentrations of 1.5-2.3 % v/v the resulting PNC size is significantly smaller than the equivalent samples made using 100 μL of precursor in solvent, at around 9 nm. PNCs prepared using solvent 3.0% v/v HBr retain an average size < 10 nm however there are occasionally particles >50 nm observed, at 3.7% v/v HBr the PNC average size increases significantly to around 32 nm. The size variation observed in the TEM images, Figure S15, is accompanied by changes in the PL spectra, Figure S16. The data reveal the impact of size and dispersity variation in the PNCs, with a progressive red-shift observed as average size increases accompanied by an increase of peak FWHM as sample dispersity increases. Overall, PNC size is reduced as precursor quantity reduces, however we did not observe any appreciable differences in the growth rate and suggest that the reduced quantity of HBr results in the formation of smaller PNCs.

***Photoconductor preparation***: Photoconductors were prepared under ambient conditions. Glass substrates were sequentially cleaned by sonicating sequentially in acetone and isopropanol, nitrogen blow-dried and treated by UV-ozone prior to PNC deposition. Following the synthesis and purification procedure PNCs stabilised by OAm in toluene were centrifuged, the supernatant removed, and 1.6 mL of toluene and 0.4 mL of MeOAc were added. Repeated centrifugation (3x5000 rpm, 5 min) resulted in a precipitate that was dispersed in 200 μL chloroform, creating a concentrated dispersion of PNCs that could be deposited as a continuous thin-film. Then, 100 μL of this solution was deposited onto a 2 × 2 cm² glass substrate and spin-coated at 1500 rpm for 20 seconds, the process repeated 4 times. Following film deposition 50 nm Au was thermally evaporated through a shadow mask (channel length of 40 μm and a width of 1000 μm).

***Material Characterisation:*** Conductivity and Dynamic light scattering (DLS) were measured by using Malvern Zetasizer Nano. Fourier Transform Infrared Spectroscopy (FTIR) was conducted on Agilent Cary 630 FTIR spectrometer with ATR sampling module. Proton nuclear magnetic resonance ($^1$H NMR) was conducted on Bruker AV-400 spectrometer using either $D_2O$ or Acetone-$d_6$ as the solvent. Transmission electron microscopy (TEM) images were collected using a JOEL 2100 PLUS at an accelerating voltage of 200 kV, samples were dispersed onto a holey carbon film on a Cu support. UV-vis absorption spectra were collected using an Agilent Cary 60 in absorption mode. Photoluminescence (PL) spectra were acquired by Agilent Cary Eclipse Fluorescence Spectrophotometer. X-ray diffraction (XRD) patterns were acquired by MPD XRD PANalytical diffractometer operated at 40kV and 40mA. X-ray photoelectron spectroscopy (XPS) was performed by Thermo Fisher K-Alpha+ equipped with a monochromated Al Ka Micro-focused X-ray source. Photoluminescence quantum yield (PLQY) and Time-resolved PL (TRPL) were acquired using an Edinburgh Instruments FLS1000 spectrometer. TRPL decays were taken with an Edinburgh Instruments EPL 375 nm picosecond pulsed diode laser. The repetition rate was controlled by an external trigger input and set to 1 MHz. The emission signal frequency was set to 3% that of the start rate to maintain single photon counting statistics. A visible PMT-980 detector was used for TRPL measurements. The photoluminescence quantum yield (PLQY) measurements were measured on the FLS 1000 fluorescence spectrophotometer with the assistance of an integrating sphere accessory (diameter = 150 nm).

***Device Characterisation***: The optoelectronic characterisations were conducted in ambient environment with a Keithley 4200 equipped with a current amplifier and the light illumination was provided by a 530 nm Thorlab LED (M530L4). Noise measurements were conducted by measuring a voltage signal and its fluctuations with a Zurich MFLI Lock-in Amplifier, amplifying with a Stanford Research System SR570 low-noise current preamplifier with a gain of 1 nA V$^{-1}$. The signal is then processed by Fast Fourier Transform (FFT) to get the noise current spectral densities.

Responsivity (R) is defined as the ratio of photogenerated current over the incident optical power output ($P_{opt}$), by considering the optical power density $E_{opt}$ and the device area (A):

$$R_c = \frac{I_{ph}}{P_{opt}} = \frac{I_{illumination} - I_{dark}}{E_{opt} A} \qquad (2)$$

The specific detectivity ($D^*$) is defined by $R_c$ and the electronic noise spectral density ($i_n \Delta f^{-1/2}$) of the device, normalising by the area of the photodetector ($A$, in this case, 0.0004 cm$^2$) to enable comparability between different photodetectors:

$$D^* = \frac{\sqrt{A \Delta f} R_C}{i_n} \qquad (3)$$

$i_n \Delta f^{-1/2}$ was extracted experimentally from the dark current recorded with a lock-in amplifier, followed by a fast Fourier transform (**Figure 6**c).

**Supporting Information**

Supporting Information is available from the Wiley Online Library.

**Acknowledgements:**

M.A.M. , E. A. and W.R.K. gratefully acknowledge the EPSRC and SFI Centre for Doctoral Training in Advanced Characterisation of Materials ( EP/S023259/1) for financial support. M.H. and M.R. thank the EPSRC (EP/T028513/1) for financial support and M.H. KAUST for baseline funding. T.J.M. thanks the Royal Commission for the Exhibition of 1851 for their financial support through a Research Fellowship and acknowledges funding from a Royal Society University Research Fellowship (URF/R1/221834).

Received: ((will be filled in by the editorial staff))

Revised: ((will be filled in by the editorial staff))

Published online: ((will be filled in by the editorial staff))

**References**

1. Mi, Bian, *et al.*, ''Real-time single-proton counting with transmissive perovskite nanocrystal scintillators'', *Nature Materials 2024*, (2024): 23, 803–809.
2. Yang, Jo, *et al.*, ''Overcoming Charge Confinement in Perovskite Nanocrystal Solar Cells'', *Advanced Materials*, (2023): 35,.
3. Dey, Ye, *et al.*, ''State of the Art and Prospects for Halide Perovskite Nanocrystals'', *ACS Nano*, (2021): 15, 10775–10981.
4. Xing, Kumar, *et al.*, ''Solution-Processed Tin-Based Perovskite for Near-Infrared Lasing'', *Advanced Materials*, (2016): 28, 8191–8196.
5. Mondal, De, Samanta, ''Achieving Near-Unity Photoluminescence Efficiency for Blue-Violet-Emitting Perovskite Nanocrystals'', *ACS Energy Letters*, (2019): 4, 32–39.
6. Schmidt, Pertegás, *et al.*, ''Nontemplate Synthesis of CH$_3$NH$_3$PbBr$_3$ Perovskite Nanoparticles'', *Journal of the American Chemical Society*, (2014): 136, 850–853.

7. Protesescu, Yakunin, *et al.*, ''Nanocrystals of Cesium Lead Halide Perovskites (CsPbX$_3$, X = Cl, Br, and I): Novel Optoelectronic Materials Showing Bright Emission with Wide Color Gamut'', *Nano Letters*, (2015): 15, 3692–3696.
8. Shamsi, Urban, *et al.*, ''Metal Halide Perovskite Nanocrystals: Synthesis, Post-Synthesis Modifications, and Their Optical Properties'', *Chemical Reviews*, (2019): 119, 3296–3348.
9. Li, Cui, *et al.*, ''Stable and large-scale organic–inorganic halide perovskite nanocrystal/polymer nanofiber films prepared via a green in situ fiber spinning chemistry method'', *Nanoscale*, (2022): 14, 11998–12006.
10. Miao, Ren, *et al.*, ''Green solvent enabled scalable processing of perovskite solar cells with high efficiency'', *Nature Sustainability*, (2023): 6, 1465–1473.
11. Lu, Tan, *et al.*, ''Green synthesis of highly stable CsPbBr$_3$ perovskite nanocrystals using natural deep eutectic solvents as solvents and surface ligands'', *Nanoscale*, (2022): 14, 17222–17229.
12. Ambroz, Xu, *et al.*, ''Room Temperature Synthesis of Phosphine-Capped Lead Bromide Perovskite Nanocrystals without Coordinating Solvents'', *Particle & Particle Systems Characterization*, (2020): 37,.
13. Hoang, Pham, *et al.*, ''A facile, environmentally friendly synthesis of strong photo-emissive methylammonium lead bromide perovskite nanocrystals enabled by ionic liquids'', *Green Chemistry*, (2020): 22, 3433–3440.
14. Chatterjee, Sen, Sen, ''Green synthesis of 3D cesium lead halide perovskite nanocrystals and 2D Ruddlesden–Popper nanoplatelets in menthol-based deep eutectic solvents'', *Materials Chemistry Frontiers*, (2023): 7, 753–764.
15. Kore, Jamshidi, Gardner, ''The impact of moisture on the stability and degradation of perovskites in solar cells'', *Cite this: Mater. Adv*, (2024): 5, 2200.
16. Cheng, Zhong, ''What Happens When Halide Perovskites Meet with Water?'', *Journal of Physical Chemistry Letters*, (2022): 13, 2281–2290.
17. Song, Lv, *et al.*, ''Surface Modification Strategy Synthesized CsPbX3 Perovskite Quantum Dots with Excellent Stability and Optical Properties in Water'', *Advanced Functional Materials*, (2023): 33,.
18. Ghinaiya, Park, Kailasa, ''Synthesis of bright blue fluorescence and water-dispersible cesium lead halide perovskite quantum dots for the selective detection of pendimethalin pesticide'', *Journal of Photochemistry and Photobiology A: Chemistry*, (2023): 444, 114980.
19. Wang, Zhang, *et al.*, ''Room temperature synthesis of CsPbX3 (X = Cl, Br, I) perovskite quantum dots by water-induced surface crystallization of glass'', *Journal of Alloys and Compounds*, (2020): 818, 152872.
20. Cheng, Yin, *et al.*, ''Water-assisted synthesis of highly stable CsPbX$_3$ perovskite quantum dots embedded in zeolite-Y †'', https://doi.org/10.1039/d0ra08311a (2021): doi:10.1039/d0ra08311a.
21. González, Mustafa, *et al.*, ''Application of natural deep eutectic solvents for the "green" extraction of vanillin from vanilla pods'', *Flavour and Fragrance Journal*, (2018): 33, 91–96.
22. Baker, Grant, ''Malic Acid Profile Active Ingredient Eligible for Minimum Risk Pesticide Use'', *https://ecommons.cornell.edu/server/api/core/bitstreams/909b05b5-9b98-48fa-8e90-8ab46bbd995a/content*, http://hdl.handle.net/1813/56132:
23. ''β-Alanine in Cell Culture'', https://www.sigmaaldrich.com/GB/en/technical-documents/technical-article/cell-culture-and-cell-culture-analysis/cell-growth-and-maintenance/beta-alanine-cell-culture:
24. Da-Wen Sun, 'Emerging Technologies for Food Processing', (Academic Press, 2005).:

25. Rosado, Duarte, Fausto, ''Vibrational spectra (FT-IR, Raman and MI-IR) of α- and β-alanine'', *Journal of Molecular Structure*, (1997): 410–411, 343–348.
26. Barańska, Kuduk-Jaworska, *et al.*, ''Vibrational spectra of racemic and enantiomeric malic acids'', *Journal of Raman Spectroscopy*, (2003): 34, 68–76.
27. Bhanavan, 'Amino Acids', *Medical Biochemistry*, (Medical Biochemistry (Fourth edition), Academic Press, 2002).:
28. Ijaz, Imran, *et al.*, ''Composition-, Size-, and Surface Functionalization-Dependent Optical Properties of Lead Bromide Perovskite Nanocrystals'', *The Journal of Physical Chemistry Letters*, (2020): 11, 2079–2085.
29. Zhang, Zhong, *et al.*, ''Brightly Luminescent and Color-Tunable Colloidal $CH_3NH_3PbX_3$ (X = Br, I, Cl) Quantum Dots: Potential Alternatives for Display Technology'', *ACS Nano*, (2015): 9, 4533–4542.
30. Zhang, Zhong, *et al.*, ''Brightly luminescent and color-tunable colloidal $CH_3NH_3PbX_3$ (X = Br, I, Cl) quantum dots: Potential alternatives for display technology'', *ACS Nano*, (2015): 9, 4533–4542.
31. Lin, Lee, *et al.*, ''Origin of Open-Circuit Voltage Enhancements in Planar Perovskite Solar Cells Induced by Addition of Bulky Organic Cations'', *Advanced Functional Materials*, (2020): 30, 1906763.
32. Elward, Chakraborty, ''Effect of Dot Size on Exciton Binding Energy and Electron–Hole Recombination Probability in CdSe Quantum Dots'', *Journal of Chemical Theory and Computation*, (2013): 9, 4351–4359.
33. Koscher, Swabeck, *et al.*, ''Essentially Trap-Free $CsPbBr_3$ Colloidal Nanocrystals by Postsynthetic Thiocyanate Surface Treatment'', *Journal of the American Chemical Society*, (2017): 139, 6566–6569.
34. Park, Lee, *et al.*, ''Surface Ligand Engineering for Efficient Perovskite Nanocrystal-Based Light-Emitting Diodes'', *ACS Applied Materials & Interfaces*, (2019): 11, 8428–8435.
35. Pan, Quan, *et al.*, ''Highly Efficient Perovskite-Quantum-Dot Light-Emitting Diodes by Surface Engineering'', *Advanced Materials*, (2016): 28, 8718–8725.
36. Lin, Lo, *et al.*, ''In situ XPS investigation of the X-ray-triggered decomposition of perovskites in ultrahigh vacuum condition'', *npj Materials Degradation*, (2021): 5, 13.
37. Pradhan, ''$CsPbBr_3$ Perovskite Nanocrystals: Linking Orthorhombic Structure to Cubic Geometry through Atomic Models and HRTEM Analysis'', *ACS Energy Letters*, (2025): 10, 1057–1061.
38. Lerum, Andersen, *et al.*, ''NMR study of the influence and interplay of water, HCl and LiCl with the extraction agent Aliquat 336 dissolved in toluene'', *Journal of Molecular Liquids*, (2020): 317, 114160.
39. Polavarapu, Nickel, *et al.*, ''Advances in Quantum-Confined Perovskite Nanocrystals for Optoelectronics'', *Advanced Energy Materials*, (2017): 7,.
40. Al-Maskari, Issac, *et al.*, ''Dye-induced photoluminescence quenching of quantum dots: role of excited state lifetime and confinement of charge carriers'', *Physical Chemistry Chemical Physics*, (2023): 25, 14126–14137.
41. Nodari, Qiao, *et al.*, ''Towards high and reliable specific detectivity in visible and infrared perovskite and organic photodiodes'', *Nature Reviews Materials*, https://doi.org/10.1038/s41578-025-00830-1 (2025): doi:10.1038/s41578-025-00830-1.
42. Maduwanthi, Jong, *et al.*, ''Stability and photocurrent enhancement of photodetectors by using core/shell structured $CsPbBr_3$/$TiO_2$ quantum dots and 2D materials'', *Nanoscale Advances*, (2024): 6, 2328–2336.
43. Wang, Yang, *et al.*, ''A high-performance photodetector based on a ZnO/$CsPbBr_3$ quantum-dot-level-contact hybrid sandwich structure'', *Journal of Materials Chemistry C*, (2025): 13, 902–909.

44. Dong, Gu, *et al.*, ''Improving All-Inorganic Perovskite Photodetectors by Preferred Orientation and Plasmonic Effect'', *Small*, (2016): 12, 5622–5632.